\documentclass[11pt]{article}
\usepackage{amsmath,amssymb,epsfig,bm}

\begin{document}

\title{\bf Cosmic deceptions due to peculiar motions}

\author{Christos G. Tsagas\\ {\small Section of Astrophysics, Astronomy and Mechanics, Department of Physics}\\ {\small Aristotle University of Thessaloniki, Thessaloniki 54124, Greece}\\ \textit{\small and}\\ {\small Clare Hall, University of Cambridge, Herschel Road, Cambridge CB3 9AL, UK}}

\date{\empty}

\maketitle

\begin{abstract}
Relative motions have long been known to mislead the unsuspecting observers to false interpretations of reality. The deceptions are usually brief and unimportant, though relative motions have also led to illusions that were both long-lasting and important. Indeed, in the history of astronomy there are several examples where relative-motion effects have misled us to gross misinterpretations. Here, we consider the possibility that our peculiar motion relative to the cosmic rest-frame can trigger deceptions on  cosmological scales. In so doing, we will demonstrate that unsuspecting observers inside bulk peculiar flows may come to the false conclusion of recent accelerated expansion, when their host universe is actually decelerating. The same observers may then erroneously attribute their apparent acceleration to an also recent dramatic change in the nature of the cosmic medium. In reality, however, nothing has really happened. Despite the appearances, the host universe keeps decelerating and its material content retains its conventional form. Nevertheless, there are ways out of these illusions. Our observers can find out that they have been deceived by their own peculiar flow, by looking for the trademark signature of relative motion in their data. This signature is nothing else but a Doppler-like anisotropy in the sky distribution of the measured deceleration parameter. To the bulk-flow observers, the universe should appear to accelerate faster along a certain point on the celestial sphere and equally slower along the antipodal. Moreover, the magnitude of the apparent dipole should decrease with increasing redshift.
\end{abstract}

\section{Introduction}\label{sI}
We have all been occasionally deceived by relative-motion effects in our everyday life. Passengers on a train, for example, may temporarily mistake their own vehicle's deceleration as acceleration of the train next to them (and vice versa). Although relative-motion deceits are typically brief, unimportant and occasionally amusing, they can also be long-lasting and quite serious. The Sun does not go around the Earth, for example, and there are no retrograde/epicyclic motions of planets, just to mention astronomical illusions that perpetuated for centuries, if not millennia.

We are moving observers in the universe. Not only on solar and galactic scales, but on cosmological scales too. Assuming that the dipole seen in the Cosmic Microwave Background (CMB) is entirely kinematic, our Local Group moves relative to the CMB frame at a speed that exceeds 600~km/sec. Moreover, there have been repeated reports of large-scale peculiar motions, the so-called bulk flows, with sizes of hundreds of Mpc, speeds of hundreds of km/sec and with our galaxy in the midst of them~\cite{WFH}. Given that observers in relative motion disagree on their interpretation of reality, it sounds plausible that they may also disagree on key features of the universe they happen to live in. Such key features could be the recent universal acceleration and the nature of the cosmic medium perceived to be responsible for it.

The theoretical possibility that the accelerated expansion of the universe is a mere illusion triggered by our peculiar motion relative to the CMB frame, was originally raised in~\cite{T1} and subsequently refined and extended in~\cite{TK}. The scenario operates within the so-called ``tilted cosmologies'', which can naturally accommodate two coordinate systems/observers in relative motion (e.g.~see~\cite{TCM}). These studies, which used linear relativistic cosmological perturbation theory, claimed that unsuspecting observers residing in slightly contracting bulk peculiar flows are likely to perceive their slower local expansion rate as acceleration of the surrounding universe. This happens because the aforementioned observers may erroneously assign negative values to the deceleration parameter measured locally in their own bulk-flow frame, while at the same time the deceleration parameter measured in the coordinate system of the universe, namely in that of the CMB, remains positive. The relative-motion effect responsible for the apparent change in the sign of the local deceleration parameter has a strong scale dependence, while its overall impact is also determined by the mean velocity of the associated bulk flow. Using data reported by several recent bulk-flow surveys~\cite{WFH}, it was found that the affected scales are large enough (between few hundred and several hundred Mpc) to make the local relative-motion effect appear as a recent global event~\cite{TK,T2}.

Unsuspecting bulk-flow observers that have been misled to believe that their universe has recently entered a phase of accelerated expansion, are likely to look for equally drastic answers in order to explain what happened. Recent dramatic changes in the nature of the cosmic medium, for example, could provide the way out of the possible deadlock. Thus, media with negative total gravitational energy (e.g.~dark energy), or even components with negative energy density (e.g.~ghost fields), could be proposed as potential answers. Nevertheless, if the same observers were to account for the effects of their peculiar motion, they would probably realise that the recent cosmic acceleration, as well as any associated changes in the nature of the material content of the universe, can be mere relative-motion illusions. Indeed, not only relative motions can trigger an apparent change in the sign of the deceleration parameter, but they can also lead to the false impression that the equation of state of the cosmic medium underwent a drastic change in its nature. Both events are illusions and they both appear to occur on the same scale/redshift, which one may call the ``transition length''. The extent of the latter depends on the scale and on the speed of the observer's bulk peculiar flow and, on using data from recent surveys, it appears to range between few hundred and several hundred Mpc~\cite{TK,T2}.

Observers suspecting that they have been deceived by their peculiar motion are likely to search for related clues in their data. The evidence in question is the trademark signature of relative motion, namely an apparent (Doppler-like) dipolar anisotropy in the sky-distribution of the deceleration parameter~\cite{T1}. To the bulk-flow observers, the universe should appear to accelerate faster towards a certain point on the celestial sphere and equally slower along the antipodal. Moreover, the magnitude of the apparent dipole should decay with increasing scale/redshft, since the peculiar velocities responsible for it are expected to do the same as well.

\section{Relative motions in cosmology}\label{sRMC}
Observers in typical galaxies, like our Milky Way, move relative to the reference frame of the universe, which has been typically identified with the coordinate system of the CMB. There, the radiation dipole is believed to vanish, which makes the CMB frame the natural coordinate system one should use to define and measure peculiar velocities in cosmology.

\subsection{The 4-velocity tilt}\label{ss4-VT}
\begin{figure}[tbp]
\centering \vspace{6cm} \includegraphics{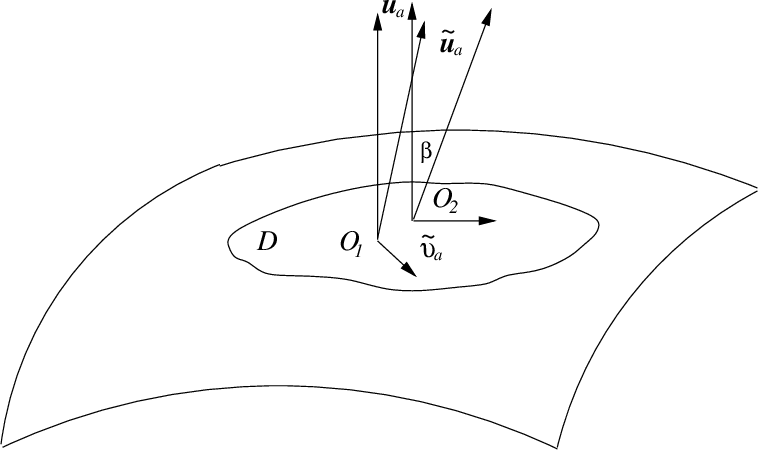} \caption{Observers ($O_1$, $O_2$) with peculiar velocity $\tilde{v}_a$ relative to the reference $u_a$-field. The 4-velocities $u_a$ and $\tilde{u}_a$ are related by the Lorentz boost $\tilde{u}_a=\gamma(u_a+\tilde{v}_a)$, where $\gamma= (1-\tilde{v}^2)^{-1/2}$. The hyperbolic ``tilt'' angle $\beta$ is defined so that $\cosh\beta=-u_a\tilde{u}^a=\gamma$, with $\gamma\simeq1$ for non-relativistic peculiar motions (i.e.~for $\tilde{v}^2\ll1$). Throughout this study the $u_a$-field is aligned with the idealised CMB frame, while $\tilde{u}_a$ is the 4-velocity of a real observer inside the bulk-flow $D$.}  \label{fig:bflow}
\end{figure}

Consider two groups of observers in an perturbed Friedmann-Robertson-Walker (FRW) universe. Also, identify one group with the ``idealised'' observers following the CMB frame and the other with the ``real'' observers moving relative to it. For non-relativistic peculiar velocities, the 4-velocity of the real observers ($\tilde{u}_a$) and that of their idealised counterparts ($u_a$) are related by
\begin{equation}
\tilde{u}_a= u_a+ \tilde{v}_a\,,  \label{4vels}
\end{equation}
where $\tilde{v}_a$ is the peculiar velocity in question. Note that $\tilde{u}_a\tilde{u}^a=-1=u_au^a, $ $u_a\tilde{v}^a=0$ by construction and $\tilde{v}^2= \tilde{v}_a\tilde{v}^a\ll1$ at the non-relativistic limit (see Fig.~\ref{fig:bflow} here and also~\cite{TCM}).

Introducing two 4-velocity fields into the spacetime implies defining two temporal directions, along $u_a$ and $\tilde{u}_a$, as well as an equal number of 3-spaces orthogonal to these 4-velocity vectors. Then, the metric ($g_{ab}$) of the host spacetime decomposes as $g_{ab}=h_{ab}+u_au_b=\tilde{h}_{ab}+ \tilde{u}_a\tilde{u}_b$, with $h_{ab}$ and $\tilde{h}_{ab}$ projecting orthogonal to their corresponding 4-velocities (i.e.~$h_{ab}u^b=0=\tilde{h}_{ab}\tilde{u}^b$). Moreover, the operators ${}^{\cdot}=u^a\nabla_a$ and ${}^{\prime}= \tilde{u}^a\nabla_a$ define time derivatives along the $u_a$ and the $\tilde{u}_a$ fields respectively, while the associated spatial gradients are ${\rm D}_a=h_a{}^b\nabla_b$ and $\tilde{\rm D}_a= \tilde{h}_a{}^b\nabla_b$. Overall, by introducing the $u_a$ and the $\tilde{u}_a$ fields, we have achieved a ``double'' 1+3~splitting of the spacetime into time and 3-D space~\cite{TCM}.

\subsection{Peculiar flux and peculiar
4-acceleration}\label{ssPFP4A}
Observers in relative motion disagree on the nature of the matter that fills the universe, even when the associated peculiar velocities are non-relativistic. Indeed, to linear order, the matter variables measured in the ``tilted'' frame of the matter and in that of the CMB are related by
\begin{equation}
\tilde{\rho}= \rho\,, \hspace{10mm} \tilde{p}= p= 0\,, \hspace{10mm} \tilde{q}_a= q_a- \rho v_a \hspace{7.5mm} {\rm and} \hspace{7.5mm} \tilde{\pi}_{ab}= \pi_{ab}= 0\,,  \label{mvars}
\end{equation}
to first approximation.\footnote{Hereafter we will always assume a pressureless cosmic medium and set both the isotropic and the viscous pressure ($p$ and $\pi_{ab}$ respectively) to zero.} Therefore, to linear order, the energy density and the pressure (both isotropic and viscous) measured in the two frames are the same. This does not apply the energy fluxes, however, which differ (i.e.~$\tilde{q}_a\neq q_a$) due to relative-motion effects alone. According to (\ref{mvars}c), even when the energy flux vanishes in one coordinate system it is nonzero in the other. Clearly, one cannot set both $q_a$ and $\tilde{q}_a$ to zero simultaneously, because then one is left with no peculiar-velocity field to study. Therefore, in the presence of peculiar motions, there is always a nonzero energy flux in the system and the cosmic medium cannot be treated as perfect~\cite{TCM}. The imperfection manifests itself as an energy flux triggered by the peculiar motion of the matter alone. It makes no real difference which frame one chooses to set the flux to zero, since it is only the relative-motion between them that matters (see~\cite{TK} for a comparison). Here, we will take the ``traditional'' approach and assume that there is no energy flux in the coordinate system of the matter (i.e.~set $\tilde{q}_a=0$), which ensures a nonzero {\it peculiar flux} ($q_a=\rho\tilde{v}_a$) in the CMB frame.

Peculiar flows are nothing else but matter in motion and moving matter means nonzero energy flux. In relativity, as opposed to Newtonian physics, energy fluxes gravitate, since they contribute to the energy-momentum tensor and through that to the local gravitational field. The purely relativistic input of the peculiar flux to the Einstein field equations feeds into the conservation laws of the energy and the momentum densities, which acquire the linear form~\cite{TCM}
\begin{equation}
\dot{\rho}= -3H\rho- {\rm D}^aq_a \hspace{15mm} {\rm and} \hspace{15mm} \rho A_a= -\dot{q}_a- 4Hq_a\,.  \label{lcls}
\end{equation}
According to the latter of the above, the nonzero peculiar flux guarantees a nonzero {\it peculiar 4-acceleration} (i.e.~$A_a\neq0$) even when the pressure is zero.\footnote{Assuming pressure-free matter and setting the flux to zero in the matter frame (recall that $\tilde{p}=p=0$ and $\tilde{q}_a=0$ to first approximation), ensures that the linear 4-acceleration vanishes there as well (i.e.~$\tilde{A}_a=0$).} This is a side effect of the fact that, in the presence of peculiar flows, the cosmic medium can no-longer be treated as perfect. Then, the relativistic momentum conservation law guarantees that imperfect media have nonzero linear 4-acceleration even when they are pressure-free~\cite{TCM}.

The form of the peculiar 4-acceleration follows from the other conservation law, namely from that for the energy density. Indeed, taking the spatial gradient of Eq.~(\ref{lcls}a), keeping in mind that $q_a=\rho\tilde{v}_a$ and confining to the linear perturbative level, we obtain~\cite{TT}
\begin{equation}
A_a= -{1\over3H}\,{\rm D}_a\tilde{\vartheta}+ {1\over3aH}\left(\dot{\Delta}_a+\mathcal{Z}_a\right)\,,  \label{lAa}
\end{equation}
where $\tilde{\vartheta}={\rm D}^a\tilde{v}_a$ is the spatial divergence of the peculiar-velocity field. Also, $\Delta_a$ and $\mathcal{Z}_a$ describe inhomogeneities in the density of the matter and in the universal expansion respectively~\cite{TCM}. The above is the linear expression one should use when studying cosmological peculiar velocities relativistically. Among others, Eq.~(\ref{lAa}) incorporates the relativistic contribution of the peculiar flux to the local gravitational field -- via the divergence ($\vartheta={\rm D}^a\tilde{v}_a$) of the peculiar velocity -- which will play the pivotal role in the outcome of this study.

\section{The deceptive role of peculiar motion}\label{sDRPM}
Relative motion effects have long been known to interfere with the way we interpret and understand our immediate surroundings, as well as the world we live in. In fact, a brief look back to the history of astronomy will reveal a number of incidents where relative-motion effects have led us to a gross misunderstanding of the way the kinematics of the universe work.

\subsection{Unsuspecting and informed bulk-flow 
observers}\label{ssUIOs}
Throughout our discussion we will distinguish between two groups of bulk-flow observers. The first are the \textit{unsuspecting observers}, namely those who are unaware of their motion relative to the cosmic rest-frame, or simply do not account for it. The second group are the \textit{informed observers}, who properly take into account the effects of their peculiar motion. This difference of perspective means that, although both groups receive the same data, they do not interpret them in the same way. More specifically, the unsuspecting observers are prone to arrive to erroneous conclusions, deceived by kinematic illusions triggered by the own peculiar flow.

In what follows we will demonstrate how the unsuspecting observers can be deceived to believe that their universe has recently entered a phase of accelerated expansion, which itself appears to be induced by a (recent as well) drastic change in the nature of the cosmic medium. At the same time, we will also demonstrate how the informed observers can find out than nothing has really happened and that everything is a mere illusion triggered by their own peculiar motion.

\subsection{Peculiar motions and the deceleration 
parameter}\label{ssPMDP}
Setting the peculiar flux to zero in the frame of the pressureless matter, ensures that the latter moves along timelike geodesics. In other words, both the peculiar flux and the peculiar 4-acceleration vanish (i.e.~$\tilde{q}_a=0=\tilde{A}_a$) in the coordinate system of the bulk-flow observers. On the other hand, $q_a$ and $A_a$ are both nonzero in the CMB frame (see \S~\ref{ssPFP4A} earlier). Therefore, to linear order the Reaychaudhuri equations read
\begin{equation}
{1\over3}\,\Theta^2q= {1\over2}\,\rho- {\rm D}^aA_a \hspace{15mm} {\rm and} \hspace{15mm} {1\over3}\tilde{\Theta}^2\tilde{q}= {1\over2}\,\tilde{\rho}\,,  \label{lRays}
\end{equation}
in the coordinate system of the CMB and of the matter respectively. This makes $q$ the deceleration parameter of the universe, whereas $\tilde{q}$ is the deceleration parameter measured locally in the bulk-flow frame. Note that $\Theta={\rm D}^au_a$ and $\tilde{\Theta}=\tilde{\rm D}^a\tilde{u}_a$ are the expansion scalars measured in the aforementioned coordinate systems, with $\tilde{\Theta}=\Theta+\tilde{\vartheta}$ to linear order and $\tilde{\vartheta}={\rm D}^a\tilde{v}_a$.

The differences between the two Raychaudhuri equations seen in (\ref{lRays}) implies that the associated deceleration parameters differ as well. Indeed, combining (\ref{lRays}a) and (\ref{lRays}b), substituting $A_a$ from Eq.~(\ref{lAa}) and assuming (for mathematical simplicity) an Einstein-de Sitter background universe, one arrives at~\cite{TK}
\begin{equation}
\tilde{q}= q+ {1\over9}\left({\lambda_H}\over\lambda\right)^2 {\tilde{\vartheta}\over H}\,,  \label{tq1}
\end{equation}
with $\lambda_H=1/H$ being the Hubble radius, $\lambda$ the scale of the bulk flow and $|\tilde{\vartheta}|/H\ll1$ throughout the linear regime. Before proceeding further, it is important to point out that, although the above expression was obtained here on an Einstein-de Sitter universe, it holds on a variety of background cosmologies. More specifically, expression (\ref{tq1}) has been shown to hold on all Friedmann and Bianchi backgrounds, provided one confines to subhorizon scales (i.e.~as long as $\lambda<\lambda_H$ -- see \cite{T2} for further discussion and details).

Following (\ref{tq1}), the deceleration parameter ($\tilde{q}$) measured by the bulk-flow observers differs form that of the actual universe ($q$) and the difference is entirely due to relatively-motion effects. Indeed, the ``correction term'' seen on the right-hand side of (\ref{tq1}) vanishes when there are no peculiar velocities (in which case $\tilde{\vartheta}={\rm D}^a\tilde{v}_a=0$ by default). This means that the difference between $\tilde{q}$ and $q$ is not real, but an apparent and deceptive relative-motion effect.

The scale dependence of the ``correction term'' in Eq.~(\ref{tq1}) is a purely relativistic effect, coming from the divergence ($\tilde{\vartheta}={\rm D}^av_a$) of the peculiar-velocity field seen on the right-hand side of (\ref{lAa}) and reflecting the gravitational contribution of the peculiar flux to the Einstein field equations. Then, despite the fact that the ratio $|\tilde{\vartheta}|/H$ is always very small at the linear level, the scale ratio $(\lambda_H/\lambda)^2$ can be considerably large on sufficiently small scales. There, the relative-motion effects can dominate and dictate the (apparent) value of the deceleration parameter ($\tilde{q}$) measured locally by the bulk-flow observers. Crucially, when the bulk flow is (slightly) contracting, that is when $\tilde{\vartheta}<0$ locally, the relative-motion effects can trigger an (apparent) change in the sign of $\tilde{q}$ from positive to negative, whereas $q$ remains positive at all times. This will deceive the unsuspecting bulk-flow observers to the illusion of accelerated expansion, despite the fact that their host universe is actually decelerating. Note that, although, the deceptive effect of relative motion is local, the affected scales are large enough (between few hundred and several hundred Mpc - see related Tables in~\cite{TK,T2}) to make it look like a recent global event.

\begin{figure}[tbp]
\centering \vspace{6cm} \includegraphics{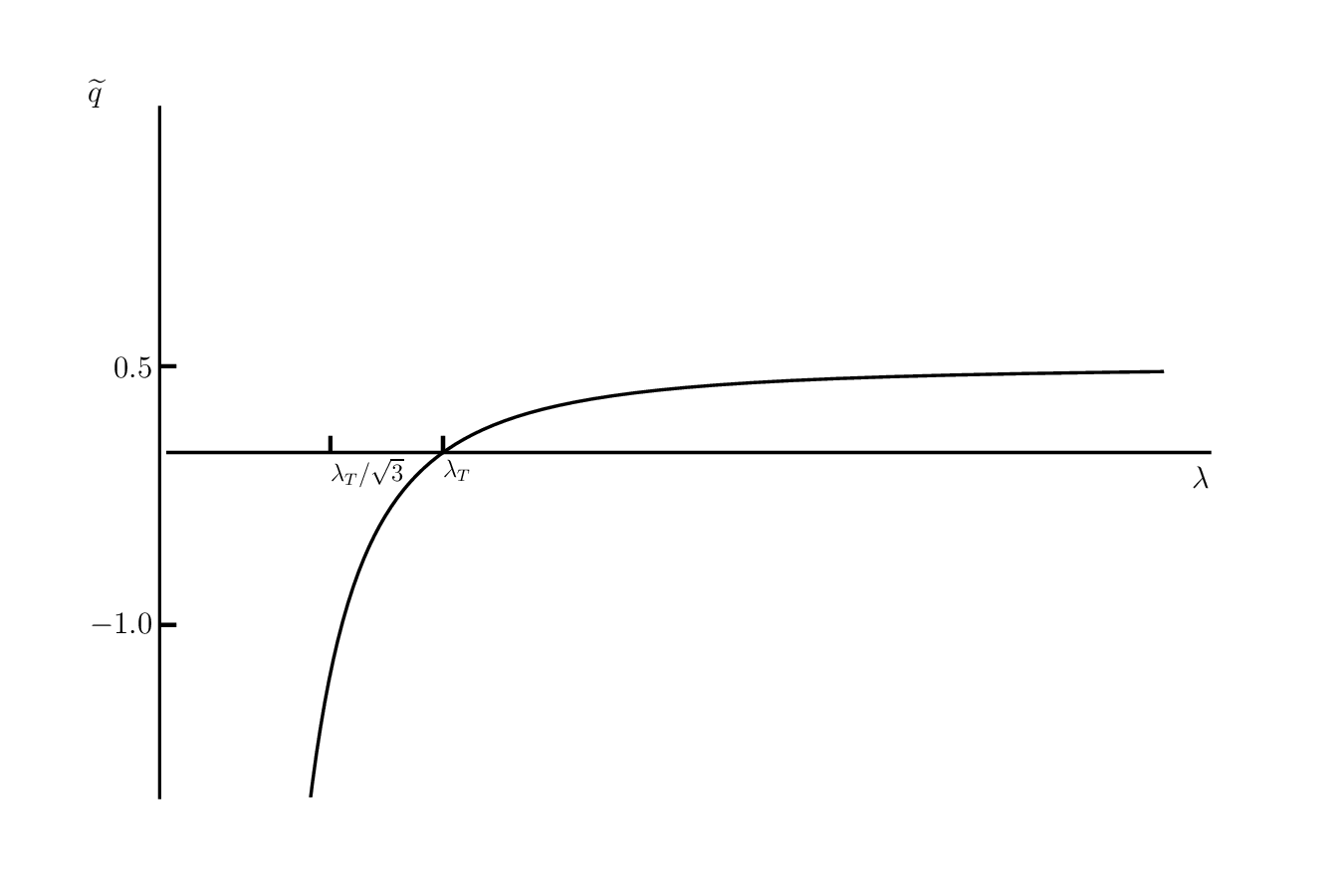} \caption{The profile of the deceleration parameter measured in the bulk-flow frame and incorporating the effects of the observers' peculiar motion (see Eq.~(\ref{tq2})). The Einstein-de Sitter value $\tilde{q}\simeq0.5$ is recovered at high redshifts where $\lambda\gg\lambda_T$. At the transition length ($\lambda_T$), Eq.~(\ref{tq2}) gives $\tilde{q}=0$ and the unsuspecting bulk-flow observers are misled to believe that the universe started to accelerate at that redshift. In fact, the acceleration rate appears to increase at progressively lower redshifts, with $\tilde{q}<-1$ when $\lambda<\lambda_T/\sqrt{3}$. To the informed bulk-flow observers, however, all this is just a deceptive side-effect of their peculiar motion relative to the cosmic rest-frame.}  \label{fig:qplot}
\end{figure}

The chances of observers finding themselves inside contracting bulk flows should be around 50\%, provided there is no natural bias in favour of contracting, or expanding, bulk flows on cosmologically relevant scales. Interestingly, a recent reconstruction of the local velocity field found its divergence to be negative, suggesting that we may actually live within a contracting bulk peculiar flow~\cite{PGC}. Confining to locally contracting bulk flows (with $\tilde{\vartheta}<0$), expression (\ref{tq1}) ensures that the sign of $\tilde{q}$ will turn negative (while $q$ is still positive) at the critical scale
\begin{equation}
\lambda_T= {1\over3}\,\sqrt{|\tilde{\vartheta}|\over qH}\,\lambda_H\,,  \label{trans}
\end{equation}
which marks the \textit{transition length} and the start of the (apparent) universal acceleration. Indeed, substituting (\ref{trans}) into the right-hand side of (\ref{tq1}), the latter recasts as
\begin{equation}
\tilde{q}= {1\over2}\,\left[1-\left({\lambda_T\over\lambda}\right)^2\right]\,,  \label{tq2}
\end{equation}
after adopting the Einstein-de Sitter value ($q=0.5$) for the deceleration parameter measured in the CMB frame. Accordingly, $\tilde{q}<q$ always and $\tilde{q}\rightarrow q$ only when $\lambda\gg\lambda_T$. Also, $\tilde{q}>0$ on scales larger than $\lambda_T$. On the other hand, on scales smaller than the transition length, the deceleration parameter measured by the bulk-flow observers ($\tilde{q}$) appears to turn negative. In fact, based on (\ref{tq2}), the scale-dependence of $\tilde{q}$ distinguishes between the following cases (see also Fig.~\ref{fig:qplot})
\begin{itemize}
\item (i) \hspace{2mm} $\tilde{q}\rightarrow0.5$, \hspace{5mm} when \hspace{5mm} $\lambda\gg\lambda_T$ \hspace{5mm} (decelerated Einstein-de Sitter expansion).
\item (ii) \hspace{2mm} $0<\tilde{q}<0.5$, \hspace{5mm} when \hspace{5mm} $\lambda_T<\lambda<\infty$ \hspace{5mm} (under-decelerated expansion).
\item (iii) \hspace{2mm} $\tilde{q}=0$, \hspace{5mm} when \hspace{5mm} $\lambda=\lambda_T$ \hspace{5mm} (coasting expansion threshold).
\item (iv) \hspace{2mm} $-1<\tilde{q}<0$, \hspace{5mm} when \hspace{5mm} $\lambda_T/\sqrt{3}<\lambda<\lambda_T$ \hspace{5mm} (accelerated expansion).
\item (v) \hspace{2mm} $\tilde{q}=-1$, \hspace{5mm} when \hspace{5mm} $\lambda=\lambda_T/\sqrt{3}$ \hspace{5mm} (de Sitter expansion threshold).
\item (vi) \hspace{2mm} $\tilde{q}<-1$, \hspace{5mm} when \hspace{5mm} $\lambda<\lambda_T/\sqrt{3}$ \hspace{5mm} (hyper-accelerated expansion).
\end{itemize}

Therefore, to the unsuspecting observers, the deceleration parameter will only seem to approach its Einstein-de Sitter value on sufficiently high redshifts, with $\lambda\gg\lambda_T$ (case (i)). Closer to the transition length, the universe will appear under-decelerated (case (ii)) and to cross the \textit{coasting expansion} threshold (where $\tilde{q}=0$) at the transition length/redshift (case (iii)). At lower redshifts the unsuspecting observes will be misled to believe the their cosmos has entered a phase of accelerated expansion (case (iv)) and even to cross the de Sitter expansion mark (when $\lambda\leq\lambda_T/\sqrt{3}$) and begin to accelerate and a progressively faster rate (cases (v) and (vi)).

On the other hand, having taken into account the effects of their peculiar motion relative to the CMB frame, the  informed observers will be able to realise that the above ``recent events'' are all mere local illusions. To them, the universe continues to decelerate close to the Einstein-de Sitter rate, at least on large cosmological scales where the linear approximation holds.

Intuitively speaking, one may think of the unsuspecting observers as passengers sitting in the back of car that drives in a broad motorway, where all the vehicles move at the same speed. If their vehicle slows down a little, without them realising it, the unsuspecting passengers are likely to believe that it is the rest of the cars that have accelerated away.

\subsection{Peculiar motions and the state of the 
matter}\label{ssPMSM}
Looking for an explanation for such a drastic change in the kinematics of their cosmos, the unsuspecting observers may consult their Raychaudhuri equation, namely the one written in the frame of the matter and given by expression (\ref{lRays}b). Since the latter applies to pressureless media and does not account for the peculiar-motion effects, negative values for $\tilde{q}$ immediately suggest $\tilde{\rho}<0$ and therefore matter fields with negative energy density. Some of the unsuspecting observers may therefore attribute their perceived universal acceleration to the recent emergence and dominance of ``ghost-like matter'' that breaks the weak energy condition (WEC). We refer the reader to~\cite{F} for a recent attempt and also to~\cite{G} for an interesting discussion.

\begin{figure}[tbp]
\centering \vspace{6cm} \includegraphics{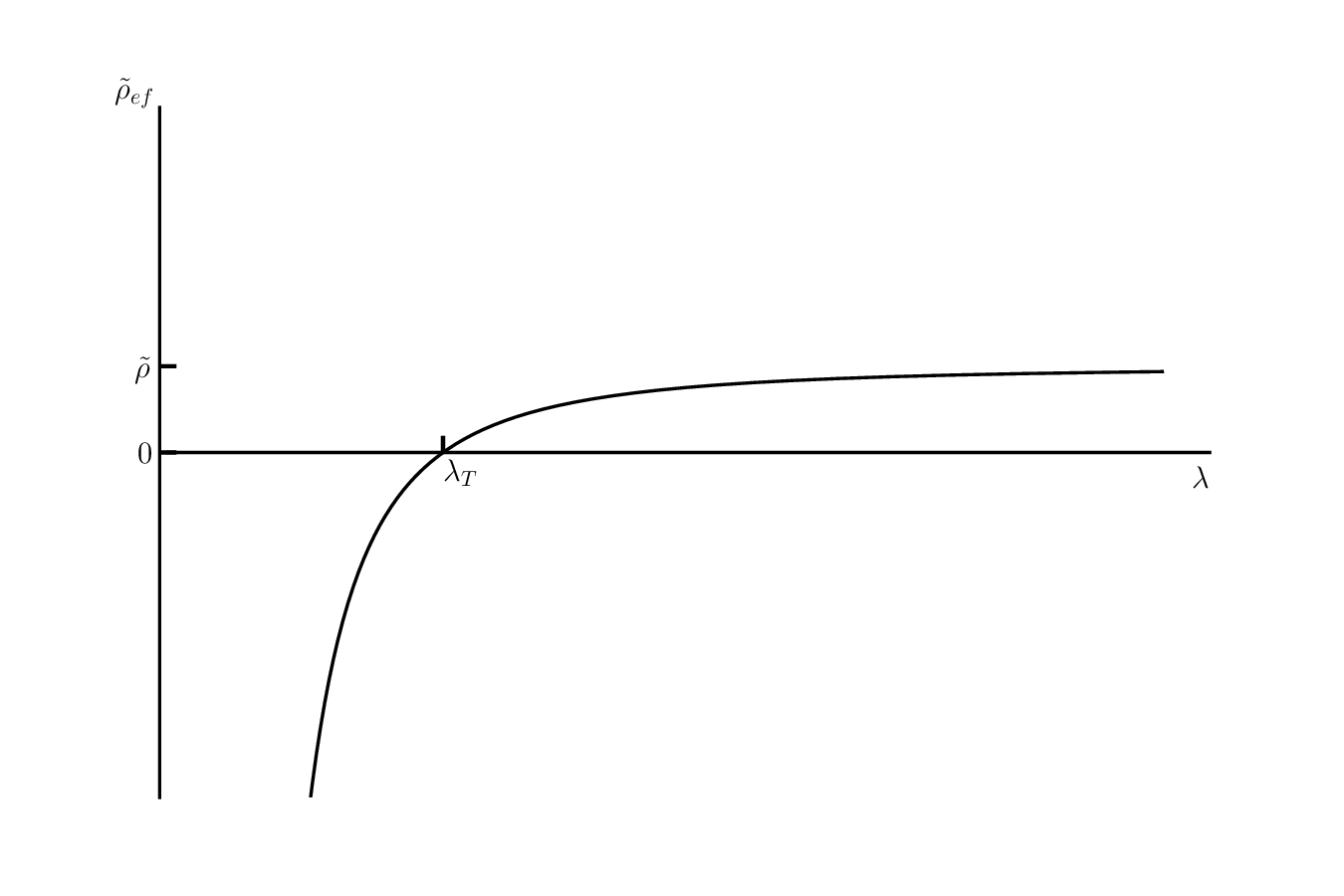} \caption{The scale/redshift profile of the effective energy density ($\tilde{\rho}_{e\hspace{-2pt}f}$), as measured in the bulk-flow frame and incorporating the effects of the observers' peculiar motion (see Eq.~(\ref{rhoef})). At relatively low redshifts, corresponding to scales smaller than the transition length ($\lambda_T$), the bulk-flow observers measure $\tilde{\rho}_{e\hspace{-2pt}f}<0$. Then, without accounting for their peculiar flow, the unsuspecting observers could be misled to believe that their universe was recently dominated by some sort of ghost-like fields with negative energy density. In contrast, to the informed bulk-flow observers, this is nothing else but a deceptive side effect of their own peculiar motion.}  \label{fig:ghost}
\end{figure}

Informed observers, who are aware of their peculiar motion and account for it, may be less hasty to embrace the ideas of their unsuspecting partners. Instead, and given the long history of deceptions triggered by relative-motion effects, these observers will soon realise that there is nothing unusual with the cosmic medium either and that everything is a mere relative-motion illusion. Indeed, being already aware that $\tilde{q}$ is not the deceleration parameter ($q$) of the actual universe, our informed observers will also know that what they measure is an effective deceleration parameter ($\tilde{q}_{e\hspace{-2pt}f}$) given by
\begin{equation}
\tilde{q}_{e\hspace{-2pt}f}= {1\over2}\,\left[1-\left({\lambda_T\over\lambda}\right)^2\right]\,,  \label{tqef}
\end{equation}
after setting $q=1/2$ in the CMB frame for the economy of the presentation. To the above measurement, the informed observers will therefore associate an effective energy density ($\tilde{\rho}_{e\hspace{-2pt}f}$) and recast their Raychaudhuri equation (see (\ref{lRays}b) earlier) into
\begin{equation}
{1\over3}\,\tilde{\Theta}^2\tilde{q}_{e\hspace{-2pt}f}= {1\over2}\,\tilde{\rho}_{e\hspace{-2pt}f}\,.  \label{lefRay}
\end{equation}
Combining relations (\ref{tqef}) and (\ref{lefRay}), while keeping in mind that $\tilde{\Theta}^2/3=\tilde{\rho}$ to linear order, leads to the following scale-dependent expression
\begin{equation}
\tilde{\rho}_{e\hspace{-2pt}f}= \left[1-\left({\lambda_T\over\lambda}\right)^2\right] \tilde{\rho}\,,  \label{rhoef}
\end{equation}
for the effective energy density associated with $\tilde{q}_{e\hspace{-2pt}f}$. Therefore, $\tilde{\rho}_{e\hspace{-2pt}f}$ is always smaller than $\tilde{\rho}$ and approaches its actual value at sufficiently high redshifts only. In other words, $\tilde{\rho}_{e\hspace{-2pt}f}< \tilde{\rho}$ on all scales and $\tilde{\rho}_{e\hspace{-2pt}f} \rightarrow\tilde{\rho}$ when $\lambda\gg\lambda_T$ only. More specifically, $\tilde{\rho}_{e\hspace{-2pt}f}$ is positive as long as $\lambda>\lambda_T$ and turns negative on scales smaller than the transition length. There, the role of the relative-motion effects mimics that of ghost matter. In this latter case, the unsuspecting observers may come to the erroneous conclusion that their cosmos entered a phase of accelerated expansion because its energy density was dominated by ghost-like fields (see Fig.~\ref{fig:ghost}).

Less radical unsuspecting observers may find the idea of ghost matter too drastic and instead suggest the dominance of ``dark energy'', namely of matter fields with nonzero pressure that break the strong energy condition (SEC) only. However, their informed counterparts will soon realise that the effects of relative motion can mimic those of dark energy as well. To demonstrate this, it suffices to introduce an effective pressure ($p_{e\hspace{-2pt}f}$) with barotropic index ($\tilde{w}_{e\hspace{-2pt}f}$) and set $\tilde{\rho}_{e\hspace{-2pt}f}=\tilde{\rho} (1+3\tilde{w}_{e\hspace{-2pt}f})$ on the right-hand side of (\ref{lefRay}). Then, on using (\ref{tqef}), we obtain the expression
\begin{equation}
\tilde{w}_{e\hspace{-2pt}f}= -{1\over3}\left({\lambda_T\over\lambda}\right)^2\,,  \label{twef2}
\end{equation}
for the effective barotropic index. The latter carries the effects of the observer's peculiar motion and has a scale dependence that distinguishes between the following cases
\begin{itemize}
\item (i) \hspace{2mm} $\tilde{w}_{e\hspace{-2pt}f}\rightarrow0$, \hspace{5mm} when \hspace{5mm} $\lambda\gg\lambda_T$ \hspace{5mm} (pressureless dust).
\item (ii) \hspace{2mm} $-1/3<\tilde{w}_{e\hspace{-2pt}f}<0$, \hspace{5mm} when \hspace{5mm} $\lambda_T<\lambda<\infty$ \hspace{5mm} (negative pressure, SEC still holds).
\item (iii) \hspace{2mm} $\tilde{w}_{e\hspace{-2pt}f}=-1/3$, \hspace{5mm} when \hspace{5mm} $\lambda=\lambda_T$ \hspace{5mm} (SEC/dark-energy divide).
\item (iv) \hspace{2mm} $-1<\tilde{w}_{e\hspace{-2pt}f}<-1/3$, \hspace{5mm} when \hspace{5mm} $\lambda_T/\sqrt{3}<\lambda<\lambda_T$\hspace{5mm} (SEC violated, dark energy).
\item (v) \hspace{2mm} $\tilde{w}_{e\hspace{-2pt}f}=-1$, \hspace{5mm} when \hspace{5mm} $\lambda=\lambda_T/\sqrt{3}$\hspace{5mm} (phantom divide/$\Lambda$-threshold).
\item (vi) \hspace{2mm} $\tilde{w}_{e\hspace{-2pt}f}<-1$, \hspace{5mm} when \hspace{5mm} $\lambda<\lambda_T/\sqrt{3}$\hspace{5mm} (phantom matter).
\end{itemize}
Accordingly, to the unsuspecting observers the cosmic medium will seem to behave as pressure-free dust only at sufficiently high redshifts, corresponding to $\lambda\gg\lambda_T$ (case (i)). Closer to the transition length, the matter will appear to acquire negative pressure, though it will still satisfy the strong energy condition (case (ii)). The \textit{dark energy} threshold (where $\tilde{w}_{e\hspace{-2pt}f}=-1/3$) is crossed at the transition length/redshift (case (iii)) and then the unsuspecting observes will be misled to believe that their cosmos has entered a phase of accelerated expansion dominated by dark energy/quintessence (case (iv)). In fact, at lower redshifts, with $\lambda=\lambda_T/\sqrt{3}$ and $\lambda<\lambda_T/\sqrt{3}$, the matter component will appear to reach and even cross the phantom divide~\cite{C} (cases (v) and (vi) respectively). Clearly, by taking into account their peculiar motion relative to the CMB frame, the  informed observers should be able to realise that all of the above are mere illusions and that the cosmic medium continues to behave as conventional pressureless dust.

\begin{figure}[tbp]
\centering \vspace{6cm} \includegraphics{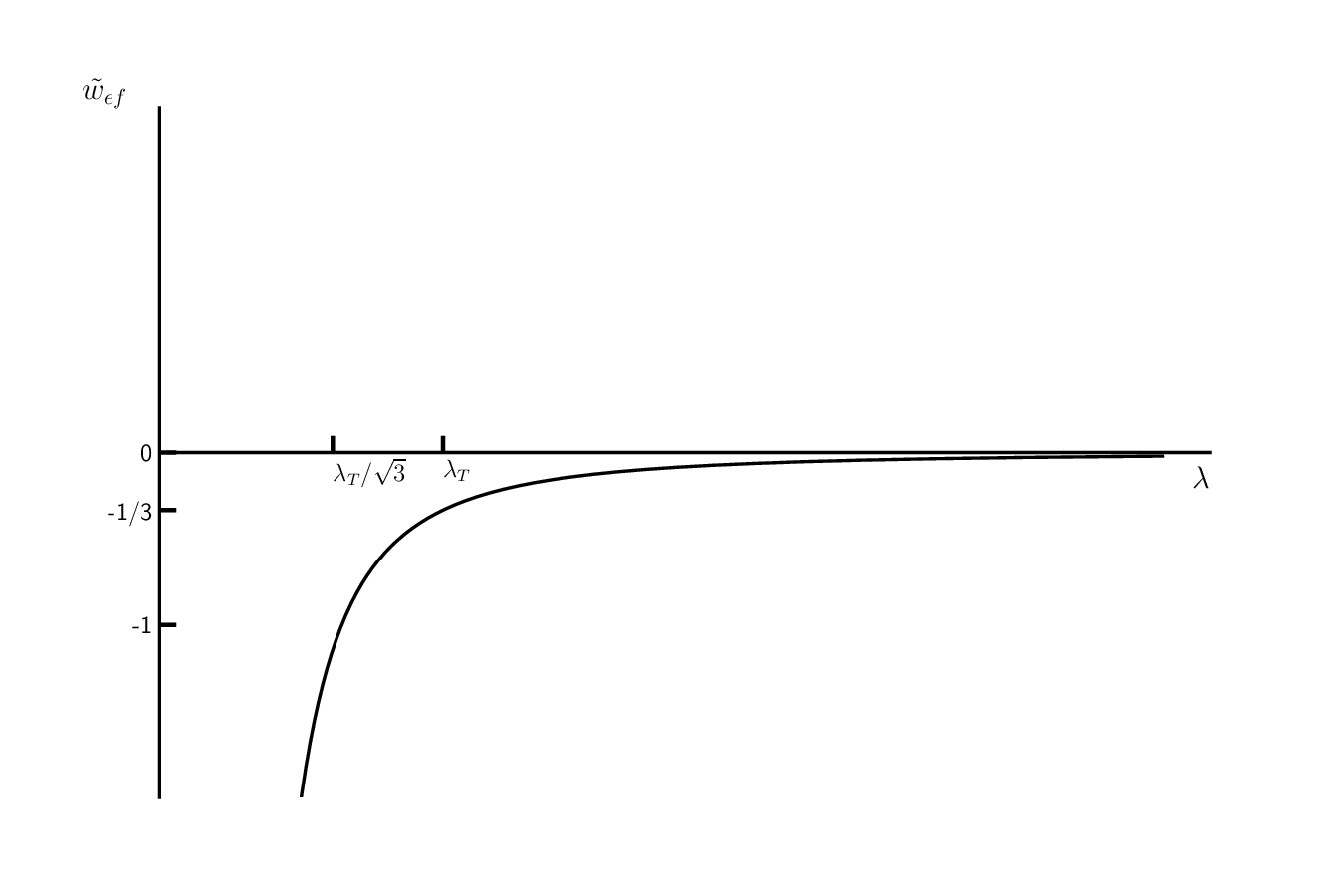} \caption{The scale/redshift profile of the effective barotropic index ($\tilde{w}_{e\hspace{-2pt}f}$), as measured in the bulk-flow frame and incorporating the effects of the observers' peculiar motion (see Eq.~(\ref{twef2})). At sufficiently high redshifts, where $\lambda\gg\lambda_T$, the bulk-flow impact is entirely negligible and the barotropic index approaches the Einstein-de Sitter value (i.e.~$\tilde{w}_{e\hspace{-2pt}f}\rightarrow0$). At lower redshifts, $\tilde{w}_{e\hspace{-2pt}f}$ turns negative, but the strong energy condition (SEC) is still satisfied as long as $\lambda>\lambda_T$. Below that threshold the SEC is violated and $\tilde{w}_{e\hspace{-2pt}f}$ can cross the dark-energy and even the phantom divides (at $\lambda=\lambda_T$ and $\lambda=\lambda_T/\sqrt{3}$ respectively). To the unsuspecting bulk-flow observers, it will therefore appear as though the universe was recently dominated by some sort of ``quintessence'', namely by a dynamically evolving dark-energy component. However, by properly accounting for their peculiar motion, the informed bulk-flow observers will soon realise that the scale/redshift evolution of $\tilde{w}_{e\hspace{-2pt}f}$ depicted above is nothing else but a mere relative-motion deception.}  \label{fig:phsntom}
\end{figure}

\section{Doppler-like signatures in the deceleration 
parameter}\label{sD-LSDP}
Relative motions are typically associated with apparent (Doppler-like) dipolar anisotropies. The CMB dipole, for example, has been attributed to  the peculiar flow of our Local Group. Similar signatures in the acceleration rate of the universe, should help the observers to tell whether the recent acceleration of their cosmos is an apparent relative-motion effect, or not.

\subsection{The deceleration tensor}\label{ssDT}
The standard deceleration parameter ($q$) is a scalar variable that monitors the mean deceleration/acceleration of the universal expansion. The latter is not fully isotropic, however, primarily due to the effects of structure formation. Here, we will describe anisotropies in the global deceleration/acceleration of the universe by introducing the deceleration tensor~\cite{T3}
\begin{equation}
Q_{ab}= -\left(h_{ab}+{9\over\Theta^2}\, h_a{}^ch_b{}^d\dot{\Theta}_{cd}\right)\,,  \label{Qab}
\end{equation}
where
\begin{equation}
\Theta_{ab}= {\rm D}_{(b}u_{a)}= {1\over3}\,\Theta h_{ab}+ \sigma_{ab}\,,  \label{Thetaab}
\end{equation}
is the expansion tensor~\cite{TCM}. Both $\Theta_{ab}$ and $Q_{ab}$ are symmetric and spacelike by construction (i.e.~$\Theta_{ab}= \Theta_{(ab)}$, $Q_{ab}=Q_{(ab)}$ and $\Theta_{ab}u^b=0=Q_{ab}u^b$). Also, the traces of (\ref{Qab}) and (\ref{Thetaab}) give
\begin{equation}
Q= 3q \hspace{15mm} {\rm and} \hspace{15mm} \Theta=3H\,, \label{Q}
\end{equation}
where $q=-[1+(3\dot{\Theta}/\Theta^2)]$ and $H=\Theta/3$ are the familiar deceleration and Hubble parameters respectively~\cite{TCM}. In other words, the deceleration tensor is a 3$\times$3 matrix with its diagonal components measuring the mean deceleration/acceleration of the expansion. The non-diagonal components of $Q_{ab}$ on the other hand, describe anisotropies.

Suppose that $n_a$ is the unit vector along a given spatial direction, so that $n_an^a=1$ and $u_an^a=0$. Then, the deceleration/acceleration of the expansion measured along $n_a$ is given by the twice contracted scalar $Q_{ab}n^an^b$. On using definition (\ref{Qab}), the latter reads
\begin{equation}
Q_{ab}n^an^b= q- {9\over\Theta^2}\,\dot{\sigma}_{ab}n^an^b\,,  \label{Qn}
\end{equation}
since $\dot{h}_{ab}n^an^b= (A_au_b+u_aA_b)n^an^b=0$~\cite{T3}. Therefore, temporal variations in the shear trigger direction-dependent deviations from the smooth universal deceleration/acceleration. Finally, when applied to an FRW universe (where $\Theta=3H$ and $\sigma_{ab}=0$ by default) expressions (\ref{Qab}) and (\ref{Qn}) reduce to~\cite{T3}
\begin{equation}
Q_{ab}= qh_{ab} \hspace{15mm} {\rm and} \hspace{15mm} Q_{ab}n^an^b= q\,,  \label{FRWQ}
\end{equation}
which is to be expected given the spatial isotropy of the Friedmann models.

\subsection{The deceleration tensors in a tilted
universe}\label{ssDTTU}
Let us consider a tilted almost-FRW universe and allow for a group of real observers living in a typical galaxy like our Milky Way and moving relative to the CMB frame with 4-velocity $\tilde{u}_a$ and bulk peculiar velocity $\tilde{v}_a$ (see Fig.~\ref{fig:bflow} in \S~\ref{ss4-VT} earlier). Written in the tilted frame of the matter, the deceleration tensor defined in (\ref{Qab}) reads
\begin{equation}
\tilde{Q}_{ab}= -\left(\tilde{h}_{ab}+{9\over\tilde{\Theta}^2}\, \tilde{h}_a{}^c\tilde{h}_b{}^d\tilde{\Theta}^{\prime}_{cd}\right)\,,  \label{tQab}
\end{equation}
with $\tilde{\Theta}_{ab}=\tilde{\rm D}_{(b}\tilde{u}_{a)}= (\tilde{\Theta}/3)\tilde{h}_{ab}+ \tilde{\sigma}_{ab}$ and the primes denoting time differentiation along the $\tilde{u}_a$-field (see \S~\ref{ss4-VT} earlier). Starting from (\ref{tQab}), employing definitions (\ref{Qab}) and (\ref{Thetaab}), while keeping up to first-order terms, we obtain~\cite{T3}
\begin{equation}
\tilde{Q}_{ab}- Q_{ab}= 2qu_{(a}\tilde{v}_{b)}- {1\over H^2} \left({\rm D}_{(b}\tilde{v}_{a)}\right)^{\cdot}\,,  \label{Qs}
\end{equation}
given that $\tilde{\rm D}_{(b}\tilde{u}_{a)}={\rm D}_{(b}u_{a)}+ {\rm D}_{(b}\tilde{v}_{a)}$ at the linear level and $\tilde{\Theta}=\Theta= 3H$ in the FRW background. The above applies to a perturbed Friedmann universe and relates the deceleration tensor ($\tilde{Q}_{ab}$) measured in the frame of the real observers to the one in the CMB frame ($Q_{ab}$) of their idealised counterparts. As expected, in the absence of peculiar motions, the two tensors coincide.

\subsection{Doppler-like dipole in the 
$\tilde{Q}$-distribution}\label{ssD-LDQ-D}
Before proceeding further, we remind the reader that we are exclusively looking for apparent dipolar anisotropies. Namely, for Doppler-like dipoles solely triggered by the peculiar motion of the real observers. Projecting expression (\ref{Qs}) along a given spatial direction (determined by the unit vector $n_a$) and using the linear commutation law $\left({\rm D}_b\tilde{v}_a\right)^{\cdot}= {\rm D}_b\dot{\tilde{v}}_a-H{\rm D}_b\tilde{v}_a$, leads to~\cite{T3}
\begin{equation}
\tilde{Q}_{ab}n^an^b- Q_{ab}n^an^b= {1\over H}\,n^a{\rm D}_a \left(\tilde{v}_bn^b\right)- {1\over H^2}\,n^a{\rm D}_a \left(\dot{\tilde{v}}_bn^b\right)\,.  \label{tQn2}
\end{equation}
The two terms on the right-hand side of the above ensure differences between the measurements made in the two frames due to relative motion effects alone. More specifically, unless these terms ``conspire'' to cancel each other out, they induce an apparent (Doppler like) dipolar anisotropy in the sky-distribution of the deceleration parameter. To demonstrate this, suppose (for simplicity) that the first term on the right-hand side of Eq.~(\ref{tQn2}) dominates. Then,
\begin{equation}
\tilde{Q}_{ab}n^an^b- Q_{ab}n^an^b= {1\over H}\,n^a{\rm D}_a \left(\tilde{v}_bn^b\right)= {1\over H}\,n^a{\rm D}_a \left(\tilde{v}\cos\phi\right)\,,  \label{tQn3}
\end{equation}
with $\phi$ being the angle between $\tilde{v}_a$ and $n_a$ (so that $0\leq\phi\leq\pi$). Without loss of generality, we may assume that there is no anisotropy due to relative-motion effects in the CMB frame and set $Q_{ab}n^an^b=q$. Then, keeping the direction vector ($n_a$) fixed, we find that~\cite{T3}
\begin{equation}
\tilde{Q}_{ab}n^an^b= q+ {1\over H}\,n^a\tilde{\rm D}_a\tilde{v}\,, \hspace{15mm} {\rm and} \hspace{15mm} \tilde{Q}_{ab}n^an^b= q- {1\over H}\,n^a\tilde{\rm D}_a\tilde{v}  \label{DtQ1}
\end{equation}
when $\tilde{v}_a\uparrow\uparrow n^a$ and $\tilde{v}_a\uparrow\downarrow n^a$ respectively (otherwise for $\phi=0$ and $\phi=\pi$). Therefore, observers moving along $n_a$ will assign the value (\ref{DtQ1}a) to the deceleration parameter measured in that direction, whereas observers moving the opposite way will assign (\ref{DtQ1}b).

\begin{figure}[tbp]
\centering \vspace{6cm} \includegraphics{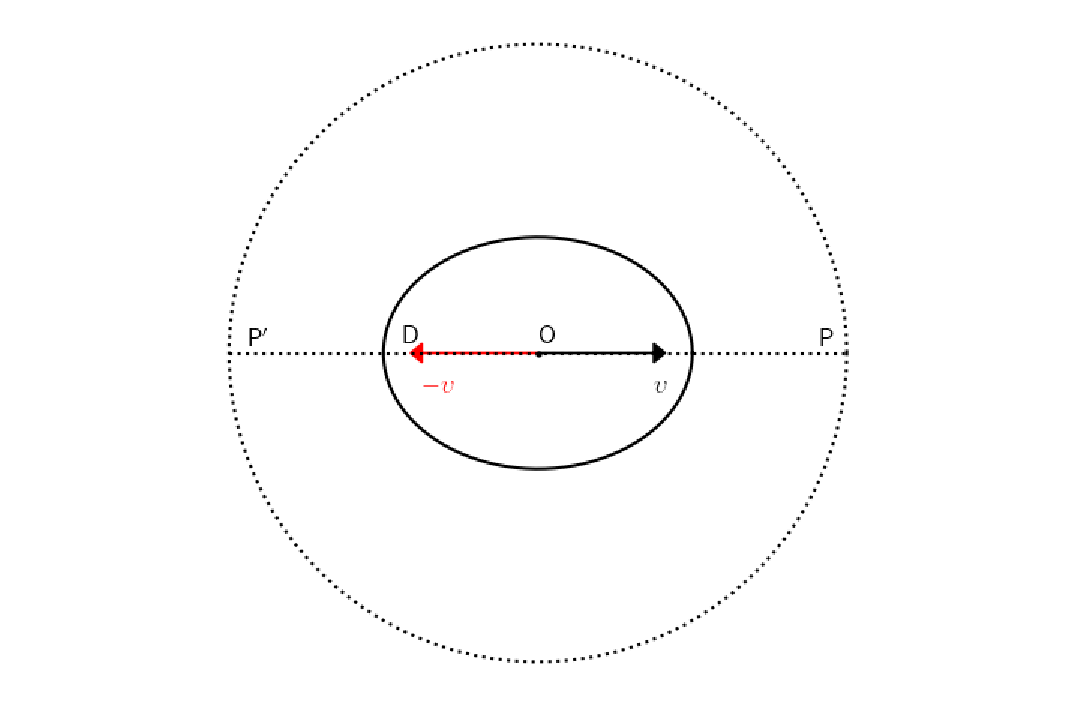} \caption{Consider the bulk flow $D$ and an isotropic distribution of identical distant (comoving) sources (dotted circle), say of supernovae type $I{\rm a}$, around the bulk-flow observer ($O$). Suppose also that the observer's peculiar velocity brings them closer to point $P$ on the supernovae distribution, while pushes them away from its antipodal $P^{\prime}$ (or vice versa). Despite the isotropy of the sources, the moving observer will ``see'' an apparent (Doppler-like) dipole in the sky distribution of the measured deceleration parameter forming along the direction of their peculiar motion (see Eqs.~(\ref{DtQ1}a) and (\ref{DtQ1}b)). Qualitatively speaking, the $\tilde{q}$-dipole is analogous to that seen in the CMB spectrum, although their axis do not necessarily coincide (see Eq.~(\ref{tQn4})).}  \label{fig:dipole}
\end{figure}

Put together, expressions (\ref{DtQ1}a) and (\ref{DtQ1}b) reveal an apparent (Doppler-like) dipolar anisotropy in the sky distribution of the deceleration parameter, as measured by the bulk-flow observers, which is solely triggered by their peculiar motion relative to the CMB frame. In particular, assuming that observers moving towards a specific point on the celestial sphere measure an increased value for their deceleration parameter, compared to its corresponding CMB value, those moving in the opposite direction (i.e.~away from the aforementioned celestial point) will measure an equally decreased value for $\tilde{Q}_{ab}n^an^b$ (and vice versa -- see Eqs.~(\ref{DtQ1}) and also Fig.~\ref{fig:dipole}). It goes without saying, that observers moving towards a given celestial  point also move away from the antipodal. Therefore, in line with (\ref{DtQ1}) the bulk-flow observers will assign  an increased/decreased value to their deceleration parameter along their peculiar motion and an equally decreased/increased value in the opposite direction (see Fig.~\ref{fig:dipole}). As a result, the sky distribution of the deceleration parameter measured by these observers should exhibit an apparent (Doppler-like) dipolar anisotropy entirely due to their peculiar motion.

Having ignored the contribution of the second term on the right-hand side of (\ref{tQn2}), the axis of the Doppler dipole lies along the direction of the peculiar motion, while its magnitude is determined by the projected gradient ($n^a\tilde{\rm D}_a\tilde{v}$) of the peculiar velocity field. In general, however, the induced anisotropy also depends on the contribution of the $\dot{\tilde{v}}_a$-term, so that~\cite{T3}
\begin{equation}
\tilde{Q}_{ab}n^an^b= q+ {1\over H}\,n^a{\rm D}_a \left[\left(\tilde{v}_b- {1\over H}\, \dot{\tilde{v}}_b\right)n^b\right]\,,  \label{tQn4}
\end{equation}
The above also ensures an apparent (Doppler-like) dipolar anisotropy in the sky distribution of the deceleration parameter, triggered by the observer's motion. This time, however, the dipole axis is not necessarily collinear with the peculiar velocity vector ($\tilde{v}_a$). In practice, this means that the dipole axis in the $q$-distribution should not necessarily coincide with that of the CMB dipole, assuming that the latter is purely kinematical and therefore collinear to $\tilde{v}_a$~\cite{T3}.

In either case, expressions (\ref{DtQ1}) and (\ref{tQn4}) suggest that the magnitude of the apparent dipole should be stronger on relatively small scales (i.e.~closer to the observer) and decay with increasing redshift (i.e.~away from the observer). This is to be expected because the responsible agent, namely the peculiar velocity field, is also expected to fade away on progressively larger scales. In addition, the magnitude of the dipole depends on the spatial gradients of both $\tilde{v}_a$ and $\dot{\tilde{v}}_a$, which also decrease with increasing scale.

Over the last decade or so, there have been a number of claims for a dipolar anisotropy in the measured distribution of the deceleration parameter. The first, to the best of our knowledge, was in the work of~\cite{CL-B}, who reported a dipole axis around 30 degrees away from that of the CMB. It was not until~\cite{CMRS}, however, that the $q$-dipole found in the JLA catalogue was attributed to our peculiar motion relative to the cosmic rest frame. An analogous dipolar anisotropy, this time in the distribution of $\Omega_{\Lambda}$, was also recently reported in~\cite{Cetal}. According to the authors, the dipole could be an artefact of our galaxy's peculiar motion, as described in~\cite{T1,TK}. These reports have received further support by a very recent analysis of the Pantheon+ data, according to which a dipolar anisotropy with a redshift decreasing magnitude does exist in the sky-distribution of the deceleration parameter~\cite{SRST,R}. Clearly, detecting a $q$-dipole with a magnitude that decays with scale/redshift will provide considerable observational support to the possibility the recent universal acceleration to be a mere illusion triggered by our galaxy's peculiar motion and thus to the \textit{tilted universe} scenario. The Pantheon+ data set and the upcoming LSST-DA survey should provide us with the opportunity

\section{Discussion}\label{sD}
The quest for an answer to the question posed by the supernovae observations of the late 1990s still goes on. While most of the cosmological community has accepted, to a larger or lesser degree, that our universe has recently entered a phase of accelerated expansion, the physical mechanism responsible for it is still unkown. Over the quarter of a century that has since past, a variety of possible answers have been proposed, all of which introduce new degrees of freedom to the existing theoretical framework. Some scenarios add new ingredients to the matter content of the universe, typically in the form of dark energy or of a cosmological constant, though more drastic suggestions (like phantom and ghost-like fields) have also appeared in the literature. Other approaches propose various changes to the gravitational theory, abandoning the Friedmann models, extra dimensions, etc. The list is, for all practical purposes, endless and it gets longer as time goes by. Also, essentially all the proposed alternatives suffer from rather excessive fine-tuning (employed to overcome the so-called ``coincidence problem'' for example) and the lack of solid physical motivation.

Here, continuing on the work of~\cite{T1,TK} we have demonstrated that our peculiar flow relative to the cosmic rest-frame can mislead the unsuspecting observer to the illusion of accelerated expansion. In fact, the unsuspecting observers may even confuse the effects of their peculiar motion with those typically attributed to dark energy/quintessence, or even to phantom and ghost-like matter. More specifically, observers that are unaware of the motion relative to the CMB frame, or simply do not account for it, will be misled to believe that their deceleration parameter turned negative and that the cosmic medium underwent a drastic change in its nature at the same redshift. All this happens at the transition length, which typically ranges between few hundred to several hundred Mpc. Therefore, the affected scales are large enough to make the unsuspecting observers treat the responsible local effects as recent global events.

In contrast, the informed observes, who are aware of their peculiar flow and also account for it, will soon realise that nothing extraordinary has happened. Instead, everything is a deception triggered by their galaxy's peculiar motion relative to the CMB frame. The universe is still decelerating and the cosmic medium is still dominated by pressureless dust (baryonic and/or not). By measuring the speed of the bulk flow they happen to live in at different redshifts, the informed observers would also be able to estimate the transition length and therefore the redshift at which the aforementioned relative-motion effects started to dominate the local kinematics. As mentioned above, in our case, the transition length varies between few hundred and several hundred Mpc.  The informed observers will also try to check whether their bulk flow is locally contracting, as required. As far as we are concerned, this is not an easy task to perform, but future data should help answer this question. Having said that, there has been a recent study claiming that our local velocity field may be actually contracting locally~\cite{PGC}.

The informed observers will also look into their data for the trademark signature of relative motion. The latter is nothing else but an apparent (Doppler-like) dipolar anisotropy in the sky-distribution of the deceleration parameter. Put another way, to the bulk-flow observers the universe should appear to accelerate faster along a certain point on the celestial sphere and equally slower in the opposite (see~\cite{T1,T3} for theoretical demonstrations). Dipolar anisotropies in the universal acceleration have been reported to exist since the work of~\cite{CL-B}, but it was only recently that they were attributed to the peculiar motion of our galaxy~\cite{CMRS,Cetal}. Moreover, if the $q$-dipole is an artefact of our peculiar flow, its magnitude must decrease with increasing scale/redshift~\cite{T3}. A dipole in the distribution of the deceleration parameter with the aforementioned decaying profile was very recently reported to exist in the Pantheon+ data~\cite{SRST,R}. If confirmed, this key prediction of the tilted universe scenario will provide strong indication that the recent universal acceleration may indeed be a mere kinematic illusion.

The suggestion that we may have been deceived, by our own peculiar motion, to believe that the universe has recently started to accelerate is a rather difficult one to accept. After all, if true, it will mean that for the last quarter of a century we have been the unsuspecting observers that mistook their own local deceleration for global acceleration of their surrounding cosmos. Be that as it may, this would not be the first time that relative-motion effects have driven us to grossly erroneous interpretations of the observations and to pointless pursuits, some of which lasted for centuries if not millennia. Moreover, in retrospect at least, the astronomical models founded on such misconceptions were sometimes bordering the absurd. Indeed, one only needs to recall how many epicycles were superimposed to ``reproduce'' the observed motions of planets and how simple everything became once the correct perspective was adopted. In this respect, the simplicity and physical plausibility of the tilted-universe scenario are significant advantages. There is no need to introduce any new physics, neither to appeal to exotic and elusive forms of matter and there is no coincidence problem either. In fact, there is no need to change anything but our perspective. We only have to account for our peculiar motion and use relativity to analyse its effects. The latter can play the role that is typically attributed to dark energy and, in so doing, create the false impression of accelerated expansion. In addition, appealing to large-scale peculiar motions has solid physical motivation. Bulk flows are not a figment of someone's wild imagination, or some speculative and farfetched idea, but an observational certainty. To the above one should add the specific predictions of the tilted model, which distinguish it from the rest of the proposed alternatives and also seem to receive observational support. On this note, we leave it to the reader to decide which scenario makes more physical and logical sense.\\

\textbf{Acknowledgements:} The author acknowledges financial support from the Hellenic Foundation for Research and Innovation (H.F.R.I.), under the ``First Call for H.F.R.I. Research Projects to support Faculty members and Researchers and the procurement of high-cost research equipment Grant'' (Project No. 789).

\end{document}